\documentclass[prd,twocolumn,showpacs,floatfix,amsmath,nofootinbib,amssymb,floatfix]{revtex4}
\usepackage{graphicx,color,dcolumn,booktabs,bm}
\usepackage{longtable,lscape}
\usepackage{txfonts}
\usepackage{overpic}
\usepackage{amssymb}
\usepackage{indentfirst}
\usepackage{feynmf}   
\usepackage{slashed}  
\usepackage{cases}
\usepackage{color}
\usepackage{multirow}
\usepackage{epstopdf}
\usepackage{graphicx,color,dcolumn,booktabs,bm}
\usepackage[colorlinks,
            citecolor=blue,
            anchorcolor=red,
            menucolor=red,
            linkcolor=red,
            filecolor=red,
            runcolor=red,
            urlcolor=blue,
            frenchlinks=red]{hyperref}

\def\lrpartial{\buildrel\leftrightarrow\over\partial}
\graphicspath{{Figures/}} %

\begin{document}

\title{Search for missing $\psi(4S)$ in the $e^+e^-\to \pi^+\pi^-\psi(2S)$ process
}
\author{Dian-Yong Chen$^{1,2}$}\email{chendy@impcas.ac.cn}
\author{Xiang Liu$^{2,3}$\footnote{Corresponding author}}\email{xiangliu@lzu.edu.cn}
\author{Takayuki Matsuki$^{4,5}$}\email{matsuki@tokyo-kasei.ac.jp}
\affiliation{$^1$Nuclear Theory Group, Institute of Modern Physics, Chinese Academy of Sciences, Lanzhou 730000, China\\
$^2$Research Center for Hadron and CSR Physics,
Lanzhou University and Institute of Modern Physics of CAS,
Lanzhou 730000, China\\
$^3$School of Physical Science and Technology, Lanzhou University,
Lanzhou 730000, China\\
$^4$Tokyo Kasei University, 1-18-1 Kaga, Itabashi, Tokyo 173-8602, Japan\\
$^5$Theoretical Research Division, Nishina Center, RIKEN, Saitama 351-0198, Japan}

\begin{abstract}
A detailed analysis to find a missing $\psi(4S)$ is made by utilizing the recent precise measurements of the cross section for the process $e^+e^-\to \psi(2S)\pi^+\pi^-$ by Belle. Assuming three resonances $Y(4360)$, $Y(4660)$, and $\psi(4S)$ to fit the data, we obtain the resonance parameters for $\psi(4S)$ as $m=4243$ MeV and $\Gamma=16 \pm 31$   MeV showing a narrow state as predicted before. A combined fit to the data $e^+ e^- \to \psi(2S) \pi^+ \pi^-, \ h_c \pi^+ \pi^-$, and $\chi_{c0} \omega$ is also performed to obtain the similar resonance parameters of $\psi(4S)$. The upper limit of the branching ratio is fitted to be $\mathcal{B}(\psi(4S)\to\psi(2S)\pi^+\pi^-) < 3\times 10^{-3}$, which can be understood by hadronic loop contributions within a reasonable range of parameters. In addition, the ratios of the branching ratios of the $\psi(4S)$ dipion transition to that of $\psi(4S)\to\chi_{c0}\omega$ are fitted, which can be further measured by BESIII and the forthcoming BelleII to confirm the existence of $\psi(4S)$.
\end{abstract}


\maketitle

\section{introduction}\label{sec1}

In 2014, the Belle Collaboration released new experimental data of the cross section of $e^+e^-\to \psi(2S)\pi^+\pi^-$ \cite{Wang:2014hta}, where two charmonium-like states $Y(4360)$ and $Y(4660)$ were confirmed again. Besides $Y(4360)$ and $Y(4660)$, the experimental data of the cross section for $e^+e^-\to \psi(2S)\pi^+\pi^-$ \cite{Wang:2014hta} show the existence of an extra narrow structure near 4.2 GeV, which inspires our interest in exploring its origin.

Since this peak is just located at the invariant mass of $\psi(2S)\pi^+\pi^-$ from the $e^+e^-$ annihilation \cite{Wang:2007ea}, we can conclude that it must have the $J^{PC}=1^{--}$ quantum number. This fact leads us to an important issue to be checked,
whether this structure around 4.2 GeV in the invariant mass of $\psi(2S)\pi^+\pi^-$ has a relation with the proposal of a missing $\psi(4S)$ in our two recent papers \cite{He:2014xna,Chen:2014sra}.

In Ref. \cite{He:2014xna}, we once predicted a missing higher charmonium $\psi(4S)$ with the mass 4263 MeV when applying the similarity between the $J/\psi$ and $\Upsilon$ families, which is supported by former theoretical calculation of the charmonium spectrum by considering the screened potential \cite{Ding:1993uy,Dong:1994zj,Li:2009zu}. Further study of the open-charm decay channels of  $\psi(4S)$ indicates that $\psi(4S)$ has very narrow decay width \cite{He:2014xna}, which is the reason why $\psi(4S)$ is still missing in the present experimental data. BESIII reported the measurement of the cross section for $e^+ e^- \to \pi^+ \pi^- h_c$ at $\sqrt{s}=3.90 \sim4.42$ GeV \cite{Ablikim:2013wzq}. Yuan performed a fit of the available experimental data of $e^+ e^- \to \pi^+ \pi^- h_c$ from 3.90 to 4.42 GeV, and found a narrow structure around 4.2 GeV, where the mass and width are reported to be $M=4216 \pm 7 $ MeV and $\Gamma=39 \pm 17$ MeV or $M=4230 \pm 10$ MeV and $\Gamma=12 \pm 36$ MeV \cite{Yuan:2013ffw}, depending on the different assumptions of the line shape trend above 4.42 GeV. The authors of Ref. \cite{He:2014xna} attributed this narrow structure existing in $e^+ e^- \to \pi^+ \pi^- h_c$ to the predicted missing $\psi(4S)$.

Later, the present authors noticed new results of $e^+e^-\to\chi_{c0}\omega$ given by BESIII  \cite{Ablikim:2014qwy}, and indicated that this observed resonance structure with mass $M=4230\pm 8$ MeV and width $\Gamma=38\pm12$ MeV in $e^+e^-\to\chi_{c0}\omega$ \cite{Ablikim:2014qwy} can be explained as the predicted $\psi(4S)$ \cite{Chen:2014sra} since the calculated branching ratio of $\psi(4S) \to \omega \chi_{c0}$ can overlap with the experimental data in a reasonable parameter range. In addition, the upper limit of a branching ratio of $\psi(4S) \to \eta J/\psi$ is also predicted to be $1.9 \times 10^{-3}$, which can be further tested by BESIII, Belle and the forthcoming BelleII\footnote{In Ref. \cite{Xu-yang:2015aya}, the branching ratio limit $B(\psi(4S) \to \eta J/\psi)<1.3\%$ was extracted by analyzing the experimental data, which is consistent with our theoretical prediction in Ref. \cite{Chen:2014sra}.}.

Besides $e^+ e^- \to \pi^+ \pi^- h_c$ \cite{Ablikim:2013wzq} and $e^+e^-\to\chi_{c0}\omega$ \cite{Ablikim:2014qwy}, we have been trying to search for any signal of the predicted $\psi(4S)$ in other hidden-charm decay channels from the $e^+e-$ annihilation. The new experimental data of $e^+e^-\to \psi(2S)\pi^+\pi^-$ \cite{Wang:2007ea} provides us good opportunity. In this work, we first fit the observed cross section for $e^+e^-\to \psi(2S)\pi^+\pi^-$ by including three resonance contributions, i.e., those of the predicted $\psi(4S)$, $Y(4360)$ and $Y(4660)$. In this way, we extract the resonance parameters of these three resonances. Comparing the obtained resonance parameters of $\psi(4S)$ with the theoretical results, we can further test the $\psi(4S)$ assignment to this structure around 4.2 GeV in the invariant mass spectrum of $\psi(2S)\pi^+\pi^-$. After fitting the experimental data of $e^+e^-\to \psi(2S)\pi^+\pi^-$ \cite{Wang:2007ea}, we
next obtain information of the branching ratio of $\psi(4S)\to \psi(2S)\pi^+\pi^-$. In this work, we also calculate the branching ratios of $\psi(4S)\to \psi(2S)\pi^+\pi^-$ and make a comparison with the extracted data, which can provide an extra test of the proposal in the present work. The detailed calculation will be given in the next section.

Since enough experimental data have been accumulated for the hidden-charm decay channels from the $e^+e^-$ annihilation, i.e., $e^+ e^- \to \pi^+ \pi^- h_c$ \cite{Ablikim:2013wzq}, $e^+e^-\to\chi_{c0}\omega$ \cite{Ablikim:2014qwy}, and $e^+e^-\to \psi(2S)\pi^+\pi^-$ \cite{Wang:2007ea}, we are able to perform a combined analysis of all these data assuming the existence of $\psi(4S)$ around 4.2 GeV. This is also the main issue in this work.


This paper is organized as follows. In the following section, a detailed analysis of fitting the $e^+e^-\to \psi(2S)\pi^+\pi^-$ data is presented on the premise that the predicted $\psi(4S)$, $Y(4360)$ and $Y(4660)$ are present.  A theoretical estimate of the branching ratio for $\psi(4S) \to \psi(2S) \pi^+ \pi^-$ is presented in Sec. \ref{sec3}. In Sec. \ref{sec4}, a combined fit to the hidden charm decay channels is performed and Sec. \ref{sec5} is devoted to a short summary.

\section{Fit of $e^+e^-\to \psi(2S)\pi^+\pi^-$ data}\label{sec2}

When looking at the cross section for $e^+ e^-\to \psi(2S) \pi^+ \pi^-$ \cite{Wang:2014hta}, one can find a number of events near 4.2 GeV other than the structures of $Y(4360)$ and $Y(4660)$.
%
%
%
Here, we perform a fit to the cross section for $e^+ e^- \to \psi(2S) \pi^+ \pi^-$ with a coherent sum of three resonances, which are $Y(4360)$, $Y(4660)$ and a resonance around $4.2$ GeV named $Y(4230)$ in this work. The total cross section can be depicted by
\begin{eqnarray}
\sigma(m) = \left| \sum_{i=0}^2 e^{i \phi_i} \mathrm{BW}_i(m) \sqrt{\frac{\mathrm{PS}_{2\to 3}(m) }{\mathrm{PS}_{2\to3}(m_i)}}   \right|^2,
\label{Eq:CS}
\end{eqnarray}
where $\phi_i$ is the phase angle between different resonances with $\phi_0=0$, and $\mathrm{PS}_{2\to 3}$ indicates the phase space of $2\to 3$ body process. The indices $i=0, 1, 2$ are assigned to the resonances $Y(4230)$, $Y(4360)$ and $Y(4660)$, respectively. The concrete form of the Breit-Wigner function of a resonance with mass $m_R$ and width $\Gamma_{R}$ is
\begin{eqnarray}
\mathrm{BW}(m) =\frac{\sqrt{12 \pi \Gamma_{R}^{e^+ e^-} \mathcal{B}(R\to f) \Gamma_{R}}}{m^2-m_R^2 +i m_R \Gamma_{R}}. \label{Eq:BWF}
\end{eqnarray}
In the Breit-Wigner function, resonance parameters and a product of a branching ratio $\Gamma^{e^+e^-}_R \mathcal{B}\left({R\to \psi(2S) \pi^+\pi^-}\right)$ are treated as free parameters.

\renewcommand{\arraystretch}{1.5}
\begin{table}[htb]
\centering
\caption{The parameters determined by fitting the experimental data of $e^+ e^- \to \psi(2S) \pi^+ \pi^-$. The symbol $f$ in the branching ratio indicates the final state, $\psi(2S) \pi^+\pi^-$. The masses and total decay widths are in units of MeV, while the product of branching ratios and the dileption decay width $\Gamma_R^{e^+ e^-} \mathcal{B}(R\to f)$ is in units of eV.  \label{Tab:psi2SpipiFit}}
\begin{tabular}{c|cccc}
\toprule[1pt]
& Sol. I & Sol. II &　Sol. III & Sol. IV\\
\midrule[1pt]
$m_{Y(4360)}$       & \multicolumn{4}{c}{$ 4356 \pm   8$}    \\
$\Gamma_{Y(4360)} $   & \multicolumn{4}{c}{ $65  \pm   10$ }  \\
$\Gamma_{Y(4360)}^{e^+e^-} \mathcal{B}(Y(4360)\to f)$  &    $ 5.8 \pm  0.8$   & $7.5 \pm 1.5$ & $6.6 \pm 2.5$ & $8.6 \pm 1.2$ \\
\midrule[1pt]
$m_{Y(4660)} $     &  \multicolumn{4}{c}{$4656 \pm 24 $}   \\
$\Gamma_{Y(4660)}$  & \multicolumn{4}{c}{$73 \pm   29 $}   \\
$\Gamma_{Y(4660)}^{e^+e^-} \mathcal{B}(Y(4660)\to f) $  &   $ 1.5  \pm 0.5$  & $1.6 \pm 0.6$ & $ 5.1 \pm 0.7$ & $5.2 \pm 0.7$ \\
$\phi_1$   &   $0.5  \pm 0.5 $  & $0.8 \pm 0.5$& $4.7 \pm 0.2$& $5.0 \pm 0.2$ \\
\midrule[1pt]
$m_{Y(4230)} $     &  \multicolumn{4}{c}{$ 4243 \pm  7 $}  \\
$\Gamma_{Y(4230)}$ &  \multicolumn{4}{c}{ $16 \pm  31$  }  \\
$\Gamma_{Y(4230)}^{ e^+e^-} \mathcal{B}(Y(4230)\to f)$    &  $ 0.5 \pm 0.5$  &$1.4 \pm 0.5$& $0.5 \pm 0.4$ & $1.4 \pm 0.5$\\
$\phi_2$   &    $ 5.5  \pm 1.0 $ & $1.7 \pm 0.5$& $5.4 \pm 1.0$& $1.6 \pm 0.5$ \\
\midrule[1pt]
$\chi^2/\mathrm{ndf}$  & \multicolumn{4}{c}{45.3/64} \\
\bottomrule[1pt]
\end{tabular}
\end{table}

\begin{figure}[t]
\centering%
\scalebox{0.6}{\includegraphics{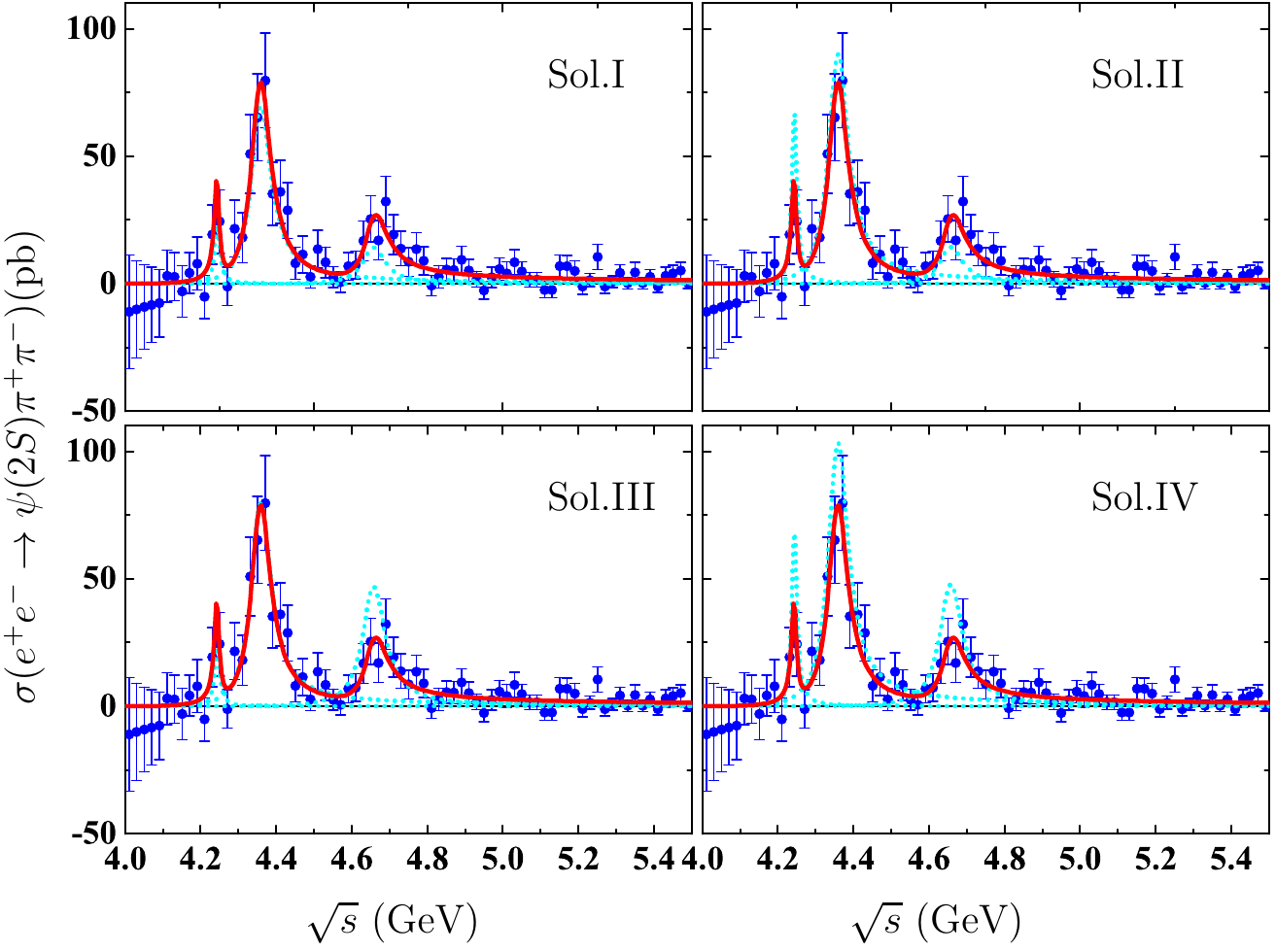}}
\caption{(color online). Our fits of different solutions to the cross section data for $Y(4230) \to \psi(2S) \pi^+ \pi^-$. The red solid, cyan dashed curves and blue dots with error bars are the total results, resonance contributions and experimental data, respectively. \label{Fig:psi2spipi-fit}}
\end{figure}

Since we include three resonances to fit the cross section with 11 free parameters, there are four independent solutions with equally good quality.
The obtained parameters are listed in Table \ref{Tab:psi2SpipiFit}. With these parameters, $\chi^2/\mathrm{ndf}$ is estimated to be $45.3/64$. The corresponding total cross sections and resonance contributions are presented in Fig. \ref{Fig:psi2spipi-fit}.
The masses and widths for $Y(4360)$ and $Y(4660)$ obtained from our fit are $m_{Y(4360)}=(4356 \pm 8)$ MeV, $\Gamma_{Y(4360)}=(65 \pm 10)$ MeV, $m_{Y(4660)} =(4656 \pm 24)$ MeV and $\Gamma_{Y(4660)}=(73 \pm 29)$ MeV, respectively. The fitted resonance parameters of $Y(4360)$ and $Y(4660)$ are consistent with those found by PDG \cite{Agashe:2014kda}. As for the resonance $Y(4230)$, the mass and width are fitted to be,
\begin{equation}\label{eq:psi4S1}
\begin{split}
m_{Y(4230)}&= 4243 \pm 7\ \mathrm{MeV}, \\
\Gamma_{Y(4230)} &= 16 \pm 31\ \mathrm{MeV}.
\end{split}
\end{equation}
Here, we notice that the resonance parameters of the new resonance $Y(4230)$ are consistent with those observed in the cross sections for $e^+ e^- \to \pi^+ \pi^- h_c $ and $e^+ e^- \to \omega \chi_{c0}$ \cite{Ablikim:2014qwy, Yuan:2013ffw}. In addition, in our previous work \cite{He:2014xna}, we have predicted a narrow charmonium state $\psi(4S)$ close to 4.2 GeV. The resonance parameters of the structure in the cross sections for $e^+ e^- \to \pi^+ \pi^- \psi(2S) $ by the Belle Collaboration, $e^+ e^- \to \pi^+ \pi^- h_c$ ($m= 4230 \pm 10\ \mathrm{MeV},\    \Gamma= 12 \pm 36 \ \mathrm{MeV}$ or $m= 4216 \pm 7\ \mathrm{MeV},\    \Gamma= 39 \pm 17 \ \mathrm{MeV}$) and $e^+e^- \to \omega \chi_{c0}$ ($m= 4230 \pm 8\ \mathrm{MeV},\    \Gamma= 38 \pm 12 \ \mathrm{MeV}$) by the BESIII Collaboration are in line with our expectation of the missing $\psi(4S)$. From these fact, we suspect that this new resonance $Y(4230)$ around 4.2 GeV in the cross sections of $e^+ e^- \to \pi^+ \pi^- \psi(2S)$ is a good candidate for our predicted $\psi(4S)$. In the next section, we will continue to test this point.

\begin{figure}[t]
\centering%
\scalebox{0.8}{\includegraphics{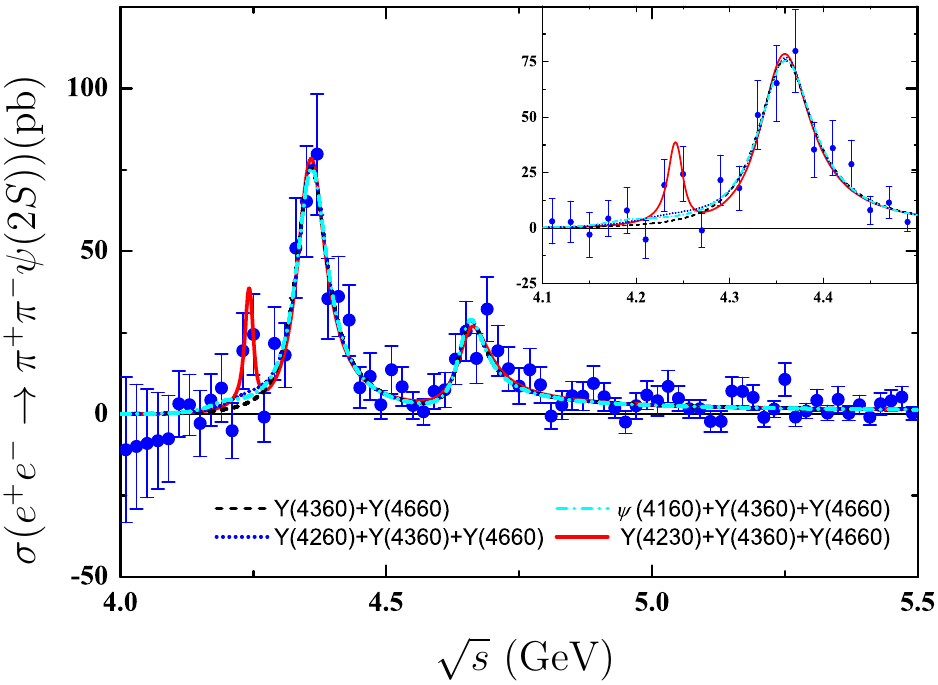}}
\caption{(color online). A comparison of the fits to the cross sections for $e^+ e^- \to \pi^+ \pi^- \psi(2S)$ with different schemes. \label{Fig:compare}}
\end{figure}

It should be noticed that in the vicinity of $4.2$ GeV, there are three resonances, $\psi(4160)$, $Y(4260)$, and the introduced $Y(4230)$ in the present work. To test the the significance of these three resonances in the cross sections for $e^+ e^- \to \pi^+ \pi^- \psi(2S)$, we fit the cross sections in four different schemes; (a) $Y(4360) + Y(4660)$, (b) $Y(4260)+Y(4360) + Y(4660)$, (c) $\psi(4160)+Y(4360) + Y(4660)$, and (d) $Y(4230)+Y(4360) + Y(4660)$. The resonance parameters of the $\psi(4160)$, $Y(4260)$, $Y(4360)$ and $Y(4660)$ are constrained by the PDG values. A comparison of the fitting curves in different schemes is presented in Fig. \ref{Fig:compare}. The dominant difference of these schemes comes from the description of the data around 4.2 GeV, where the fitting curves are presented in the subfigures of Fig. \ref{Fig:compare}. The $\chi^{2}$ for the different schemes are $51.2,\ 48.8,\ 49.2$, and $45.3$, respectively, which indicates that the experimental data in the vicinity of $4.2$ GeV of the cross sections for $e^+ e^- \to \pi^+ \pi^- \psi(2S)$ favor a narrow structure $Y(4230)$.

Besides the resonance parameters of $Y(4230)$, we also obtain the product of the branching ratio to $e^+e^-\psi(2S)$ and the $e^+e^-$ partial width (abbreviated as $\Gamma\mathcal{B}$), which can provide us with further more information about this state and is presented in Table \ref{Tab:psi2SpipiFit}. Since $Y(4230)$ is a narrow state and far away from $Y(4660)$, interference between $Y(4230)$ and $Y(4660)$ is very weak and ignorable, while interference between $Y(4230)$ and $Y(4360)$ is significant. $\Gamma\mathcal{B}$ can be divided into two groups according to interference between $Y(4230)$ and $Y(4360)$. For Sols. I and III, this interference is constructive. The center value of $\Gamma\mathcal{B}$ in these two solutions is $0.5$ eV, which is comparable to zero when including the errors. For Sols. II and IV, this interference is destructive, in which case $\Gamma\mathcal{B}$ is $(1.4 \pm 0.5)$ eV. From our fit, we obtain the upper limit of $\Gamma\mathcal{B}$ to be $1.9 $ eV.

\section{Meson loop contributions to $\psi(4S) \to \psi(2S) \pi^+ \pi^-$} \label{sec3}

\begin{figure}[htb]
\centering%
\scalebox{0.4}{\includegraphics{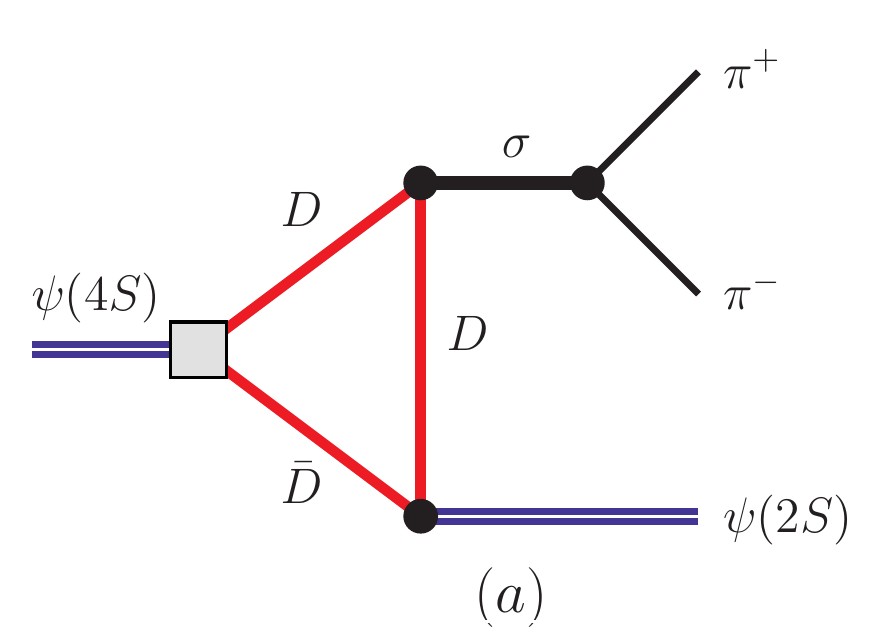}}
\scalebox{0.4}{\includegraphics{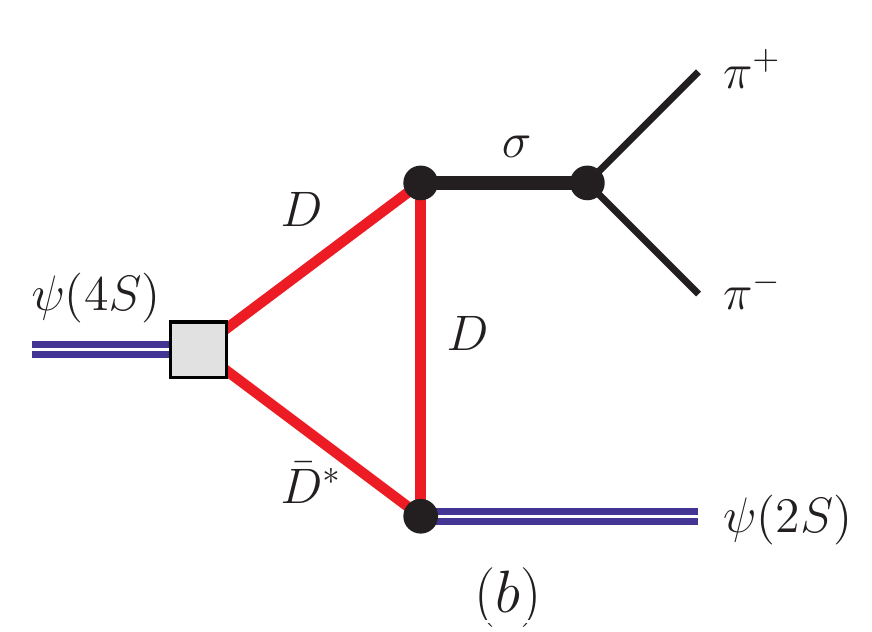}}
\scalebox{0.4}{\includegraphics{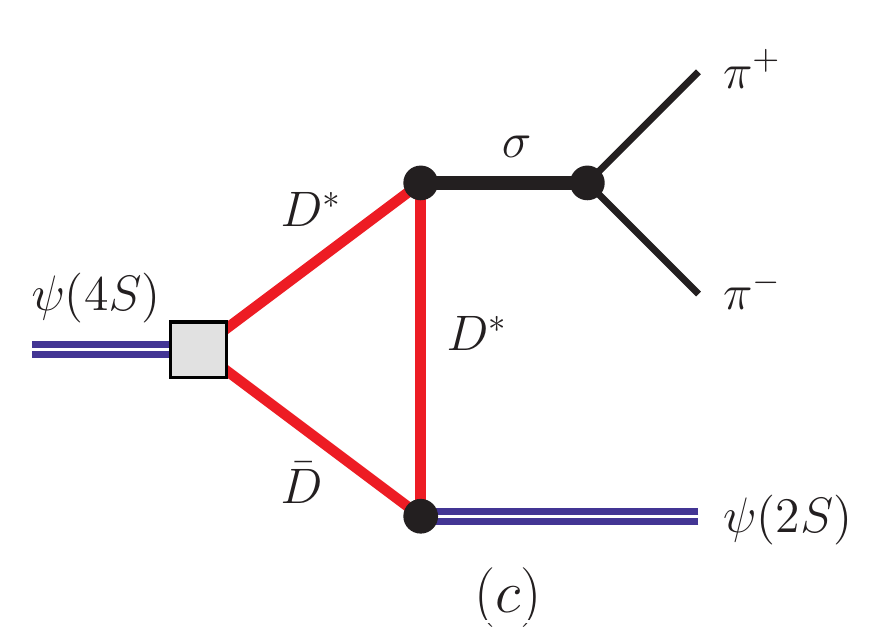}}
\scalebox{0.4}{\includegraphics{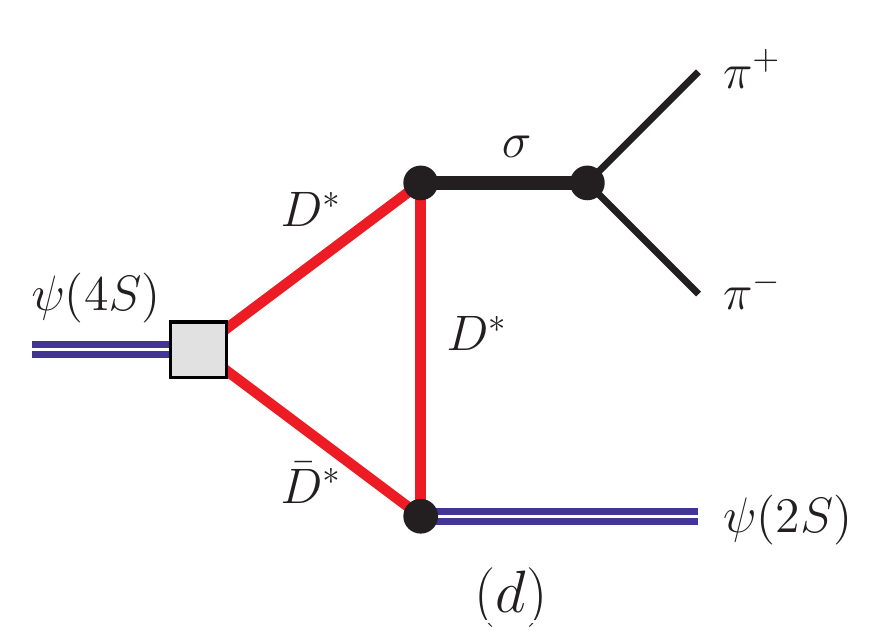}}
\caption{(color online). Typical meson loop contributions to $\psi(4S) \to \psi(2S) \pi^+ \pi^-$, where the dipion comes from $\sigma$ meson. \label{Fig:Tri}}
\end{figure}

From our fit to the experimental data of $e^+ e^- \to \psi(2S) \pi^+ \pi^-$, we find that the product of the branching ratio to $\pi^+\pi^-\psi(2S)$ and the $e^+e^-$ partial width, $\Gamma_{Y(4230)}^{e^+e^-} \mathcal{B}(Y(4230) \to \psi(2S) \pi^+ \pi^-)$ ($\Gamma\mathcal{B}$), is dependent on interference between $Y(4230)$ and $Y(4360)$. If interference is constructive, $\Gamma\mathcal{B}$ is comparable to zero, whereas if interference is destructive, $\Gamma\mathcal{B}$ is $1.4 \pm  0.5$ eV, which indicates that the upper limit of $\Gamma\mathcal{B}$ is $1.9$ eV.
As we have discussed in the previous section, the resonance parameters of $Y(4230)$ are consistent with the expectations of the missing $\psi(4S)$. To further check the possibility of $Y(4230)$ as the $\psi(4S)$, we need to understand the fitted product of the branching ratio to $\pi^+\pi^-\psi(2S)$ and the $e^+e^-$ partial width, which is less than 1.9 eV. Because the measurement of the $\psi(4S)$ dilepton decay width is not available at the moment, we adopt the theoretical estimate of $\Gamma^{e^+ e^-}_{\psi(4S)}$, which is 0.63 keV \cite{Dong:1994zj} or 0.66 keV \cite{Li:2009zu} in the quark potential model.

In Ref. \cite{Dong:1994zj}, authors interpreted the observed $\psi(4160)$ as a $\psi(4S)$ state and the $\psi(4415)$ as a $\psi(5S)$ state to solve the contradiction observed in dilepton decay widths between experimental and theoretical data. However, with this assignment, the $\psi(2D)$ state would not have any experimental counterpart. We notice that the theoretical mass of $\psi(4S)$ was estimated to be $4274$ MeV \cite{Dong:1994zj}, which is about 120 MeV higher than the mass of $\psi(4160)$ and is comparable with the $Y(4230)$ discussed in this work. Thus, the $\psi(4S)$ predicted by the screened potential model in Ref. \cite{Dong:1994zj} is more likely to be $Y(4230)$ existing in $e^+ e^- \to \pi^+ \pi^- \psi(2S)$ rather than $\psi(4160)$.

With the theoretical value of $\Gamma_{\psi(4S)}^{e^+ e^-}$, we can roughly estimate the upper limit of the branching ratio for $\psi(4S) \to \psi(2S) \pi^+ \pi^-$ to be $3.0 \times 10^{-3}$ with the assumption of $Y(4230)$ as the missing $\psi(4S)$. The interpretation on this branching ratio in the $\psi(4S)$ framework can further test the nature of this state.

Here, one should specify that the mass of the missing $\psi(4S)$ is above the threshold of a pair of charmed mesons, which means the $\psi(4S)$ dominantly decays into a charmed meson pair \cite{He:2014xna}. A charmed meson pair can transit into a charmonium and a light meson in the final state, this mechanism, named the meson loop mechanism, plays a crucial role in understanding the hidden charm/bottom decay behaviors of a higher heavy quarkonium \cite{Chen:2014sra, Chen:2013cpa, Chen:2014ccr, Li:2007au, Zhang:2009kr, Guo:2009wr}. The process $\psi(4S)$ decaying into $\psi(2S) \pi^+ \pi^-$ occurs via a charmed meson loop as shown in Fig. \ref{Fig:Tri}. Here, we adopt the effective Lagrangian approach to estimate meson loop contributions to the decay $\psi(4S) \to \psi(2S) \pi^+ \pi^-$. The effective interaction between charmonium and a charmed meson pair can be constructed in the heavy quark limit \cite{Casalbuoni:1996pg, Casalbuoni:1992fd, Colangelo:2003sa}. The specific effective Lagrangians for $\psi \mathcal{D}^{(\ast)} \mathcal{D}^{(\ast)\dagger}$ are
\begin{eqnarray}
\mathcal{L}_{\psi \mathcal{D}^{(\ast)} \mathcal{D}^{(\ast)}}
&=& -ig_{\psi \mathcal{DD} } \psi_\mu (\partial^\mu
\mathcal{D} \mathcal{D}^\dagger- \mathcal{D}
\partial^\mu \mathcal{D}^\dagger) \nonumber\\
&& + g_{\psi
\mathcal{D}^\ast \mathcal{D}} \varepsilon^{\mu \nu \alpha \beta}
\partial_\mu \psi_\nu (\mathcal{D}^\ast_\alpha \lrpartial_\beta
\mathcal{D}^\dagger -\mathcal{D} \lrpartial_\beta
\mathcal{D}_\alpha^{\ast \dagger} ) \nonumber\\
&& + ig_{\psi
\mathcal{D}^\ast \mathcal{D}^\ast} \psi^\mu
(\mathcal{D}^\ast_\nu \partial^\nu \mathcal{D}^{\ast \dagger}_\mu
-\partial^\nu \mathcal{D}^{\ast}_\mu \mathcal{D}^{\ast \dagger}_\nu
\nonumber\\&&-\mathcal{D}^\ast_\nu \lrpartial_\mu \mathcal{D}^{\ast \nu
\dagger}), \label{Eq:Lag1}
\end{eqnarray}
where $\mathcal{D}^{(\ast)}=(D^{0(\ast)},D^{+(\ast)}, D_s^{+(\ast)})$. The effective coupling constants between $\psi(2S)$ and the charmed meson pair can be related to the decay constant of $\psi(2S)$ by $g_{\psi(2S) DD}=  g_{\psi(2S) D^\ast D^\ast} {m_{D^\ast}}/{m_D} =g_{\psi(2S) D^\ast D} m_{\psi} \sqrt{m_D/m_{D^\ast}} =m_{\psi(2S)} /f_{\psi(2S)} $. The $\psi(2S)$ decay constant $f_{\psi(2S)}$  can be evaluated from the leptonic decay width of $\psi(2S)$.  With $\Gamma_{\psi(2S)}^{e^+ e^-} =2.36$ keV, we have $f_{\psi(2S)} =297 $ MeV. As for the interaction between $\psi(4S)$ and charmed meson pairs, we adopt the same formula as shown above for $\psi(2S)$, but the related coupling constants are evaluated by the corresponding decay widths. Since the experimental measurements for the open charm decay of $\psi(4S)$ are not available yet, here we adopt the theoretical predictions based on the quark pair creation model \cite{He:2014xna}. As shown in Ref. \cite{He:2014xna}, the open charm decay width depends on the parameter $R$ introduced in the spatial wave function of $\psi(4S)$. Thus, the coupling constants estimated from the partial decay width will depend on the parameter $R$. The values of the coupling constants related to $\psi(4S)$ have been given in our previous work \cite{Chen:2014sra}.

The effective Lagrangians related to the $\sigma$ meson are \cite{Meng:2007cx, Chen:2011xk, Chen:2011zv, Chen:2011qx},
\begin{eqnarray}
\mathcal{L}_{\sigma \mathcal{D}^{(\ast)} \mathcal{D}^{(\ast)}} &=&-g_{\sigma \mathcal{D} \mathcal{D}} \mathcal{D} \mathcal{D}^\dagger \sigma + g_{\sigma \mathcal{D}^\ast \mathcal{D}^\ast } \mathcal{D}^\ast \mathcal{D}^{\ast \dagger} \sigma,\nonumber\\
\mathcal{L}_{\sigma \pi \pi} &=& g_{\sigma \pi \pi} \sigma \pi \pi,
\end{eqnarray}
where $g_{\sigma DD}= g_{\sigma D^\ast D^\ast} =m_{D^\ast} g_{\pi}/\sqrt{6}$ with $g_{\pi}=3.73$ \cite{Bardeen:2003kt, Liu:2008xz}. The coupling of $\sigma \pi \pi$ is evaluated by the decay width of the $\sigma$ meson, where, $\Gamma_{\sigma}^{\mathrm{Tot}} \simeq \Gamma_{\sigma\to \pi^+ \pi^-} + \Gamma_{\sigma \to \pi^0 \pi^0}/2$. In the present work, we take $m_{\sigma}= 526$ MeV and $\Gamma_{\sigma} =302$ MeV \cite{Aitala:2000xu}.

In the triangle diagram, the exchanged charmed mesons are off-shell. To describe the off shell effect and the structure of the exchange mesons and to regularize the divergence, we introduce a form factor in the amplitudes. In the present work, we adopt a monopole form form factor as \cite{Chen:2014sra, Chen:2014ccr, Chen:2013cpa},
\begin{eqnarray}
\mathcal{F}(q^2,\Lambda^2) =\frac{m_E^2-\Lambda^2}{q^2-\Lambda^2},
\end{eqnarray}
where $q$ and $m_E$ are the momentum and the mass of the exchanged meson, respectively. The parameter $\Lambda$ can be reparametrized as $\Lambda= m_E+\alpha_{\Lambda} \Lambda_{QCD}$ with $\Lambda_{QCD}=220$ MeV. The dimensionless parameter $\alpha_{\Lambda}$ is of order 1 and dependent on the specific process \cite{Cheng:2004ru}.

With the above preparations, we parametrize the amplitude of $\psi(4S)(p_0) \to \psi(2S)(p_1) \pi^+(p_2) \pi^-(p_3)$ in the form,
\begin{eqnarray}
\mathcal{M} &=& \epsilon_{\psi(4S)}^\mu \epsilon_{\psi(2S)}^\nu (f_S g_{\mu \nu} +f_{D} p_{1\mu} p_{0\nu})
\nonumber \\
&&\times\frac{g_{\sigma \pi \pi}}{(p_2+p_3)^2-m_{\sigma}^2 +im_\sigma\Gamma_{\sigma}(m_{\pi\pi})},
\end{eqnarray}
where $f_S$ and $f_D$ are evaluated from the loop integral of the triangle diagrams listed in Fig. \ref{Fig:Tri}. For the broad resonance $\sigma$, we introduce a momentum-dependent decay width as \cite{Aitala:2000xu}
\begin{eqnarray}
\Gamma_{\sigma}(m_{\pi\pi}) = \Gamma_{\sigma} \frac{m_\sigma}{m_{\pi\pi}} \frac{\left|\vec{p}(m_{\pi\pi})\right|}{\left|\vec{p}(m_\sigma)\right|}
\end{eqnarray}
where $|\vec{p}(m_{\pi\pi})| =\sqrt{m_{\pi\pi}^2/4-m_\pi^2}$ is the pion momentum in the mother particle rest frame. With the above formula, we get the differential partial decay width as
\begin{eqnarray}
d\Gamma = \frac{1}{(2 \pi)^3} \frac{1}{32 m_{\psi(4S)}^2} \overline{|\mathcal{M}|^2} dm^2_{\psi(2S) \pi} dm^2_{\pi\pi},
\end{eqnarray}
where the overline above $|\mathcal{M}|^2$ indicates the average over the spin of initial $\psi(4S)$. We can estimate the partial decay width by integrating over $m_{\psi(2S) \pi}$ and $m_{\pi\pi}$.

\begin{figure}[htb]
\centering%
\scalebox{0.6}{\includegraphics{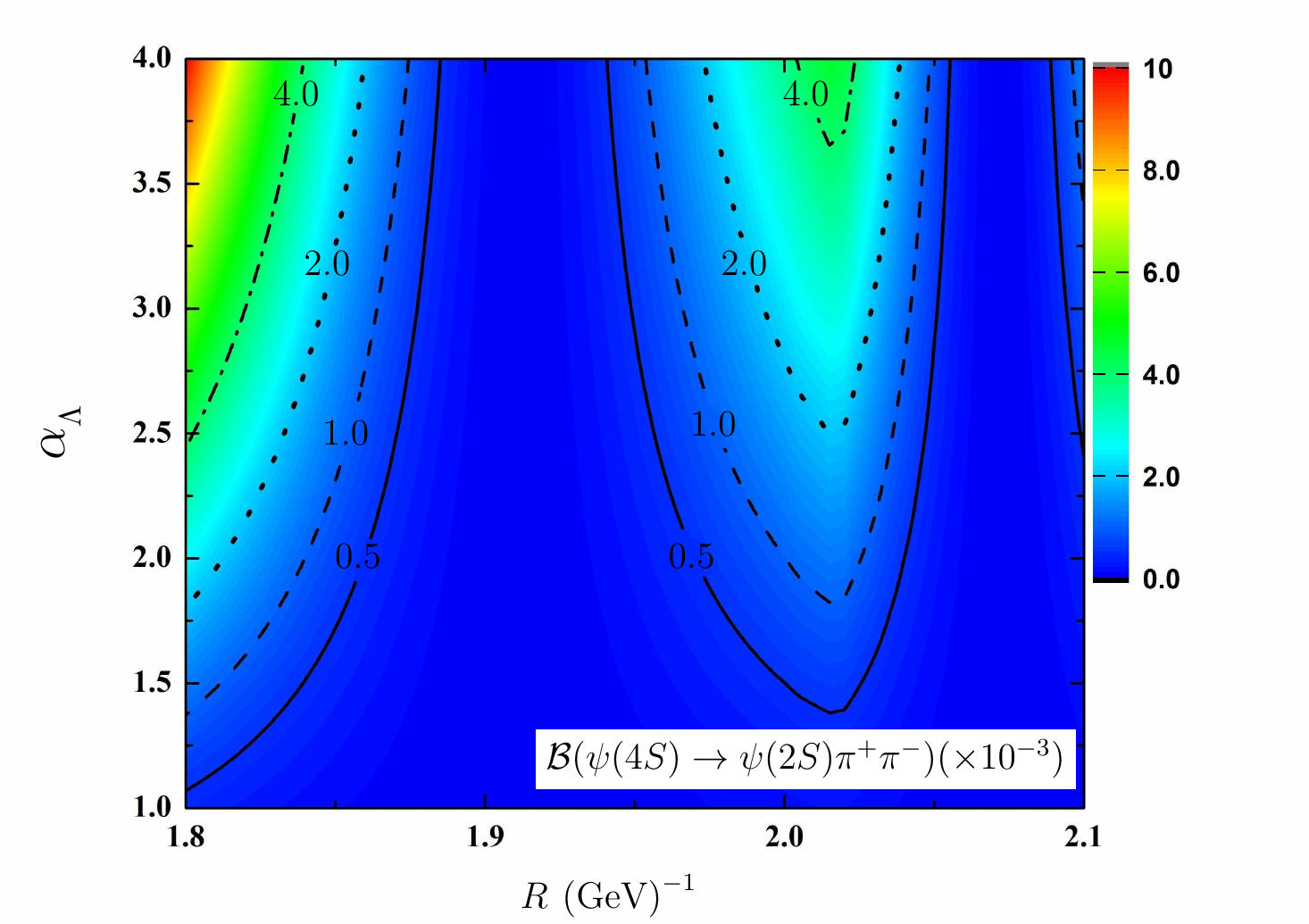}}
\caption{(color online). The $R$ and $\alpha_{\Lambda}$ dependence of the branching ratio for $\psi(4S) \to \psi(2S) \pi^+ \pi^-$. \label{Fig:DWML}}
\end{figure}

The meson loop contributions to the branching ratio for $\psi(4S) \to \psi(2S) \pi^+ \pi^-$ depend on both parameters $R$ and $\alpha_{\Lambda}$, which are introduced in the wave function of $\psi(4S)$ and the form factors in the amplitudes. The $R$ and $\alpha_{\Lambda}$ dependence of the branching ratio for $\psi(4S) \to \psi(2S) \pi^+ \pi^-$ is presented in Fig. \ref{Fig:DWML}. Some contour lines with several typical values of the branching ratio are also shown. We find that the branching ratio $\mathcal{B}(\psi(4S) \to \psi(2S) \pi^+ \pi^-)$ resulting from meson loop contributions overlaps with the upper limit, $3.0\times 10^{-3}$, obtained by fitting  the cross section for $e^+ e^- \to \psi(2S) \pi^+ \pi^-$. This fact further supports the assignment of  $Y(4230)$ existing in the $e^+ e^- \to \psi(2S) \pi^+ \pi^-$ process as the missing $\psi(4S)$.

\section{Combined fit to $e^+ e^- \to \psi(2S) \pi^+ \pi^-, \ h_c \pi^+ \pi^-, \ \chi_{c0} \omega$} \label{sec4}

As we have indicated in Section \ref{sec2}, the resonance parameters of the new structure, $Y(4230)$, obtained by fitting the experimental data of $e^+ e^- \to \psi(2S) \pi^+ \pi^-$ are consistent with those obtained by fitting the cross sections of $e^+ e^- \to h_c \pi^+ \pi^-$ and $e^+ e^- \to \chi_{c0} \omega$, which indicates that the structures in these three processes may come from the same source. To further test this conjecture, we shall perform a combined fit to the experimental data of the cross sections for $e^+ e^- \to \psi(2S) \pi^+ \pi^-$, $e^+ e^- \to h_c \pi^+ \pi^-$ and $e^+ e^- \to \chi_{c0} \omega$ simultaneously.

We adopt the same formula as Eq. (\ref{Eq:CS}) to describe the cross section for $e^+ e^- \to \chi_{c0} \omega$ by replacing the $2\to 3$ phase space with a $2\to 2$ phase space. As for $e^+ e^- \to h_c \pi^+ \pi^-$, we notice that the experimental data for the cross section of $e^+ e^- \to h_c \pi^+ \pi^-$ above 4.4 GeV are not available at present \cite{Yuan:2013ffw}. Here, we adopt two schemes to describe the cross section for $e^+ e^- \to h_c \pi^+ \pi^-$ depending on the tendency above 4.4 GeV \cite{Yuan:2013ffw}. In the first scheme, the cross section for $e^+ e^- \to h_c \pi^+ \pi^-$ goes down above 4.4 GeV, where two Breit-Wigner functions are adopted to depict the cross section as Eq. (\ref{Eq:CS}). In the second scheme, the cross section goes up above 4.4 GeV, where the cross section is described by the coherent sum of a Breit-Wigner function and the $2\to 3$ phase space in the form,
\begin{eqnarray}
\sigma_{e^+ e^- \to h_c \pi^+ \pi^-} = \left| c \sqrt{\mathrm{PS}_{2\to 3}(m)} +e^{i\phi} \mathrm{BW}(m) \sqrt{\mathrm{PS}_{2\to3}(m)\over \mathrm{PS}_{2\to3}(m_R)}\right|^2. \nonumber\\
\label{Eq:hcII}
\end{eqnarray}

\subsection{Scheme I}


\begin{figure}[htb]
\centering%
\scalebox{0.63}{\includegraphics{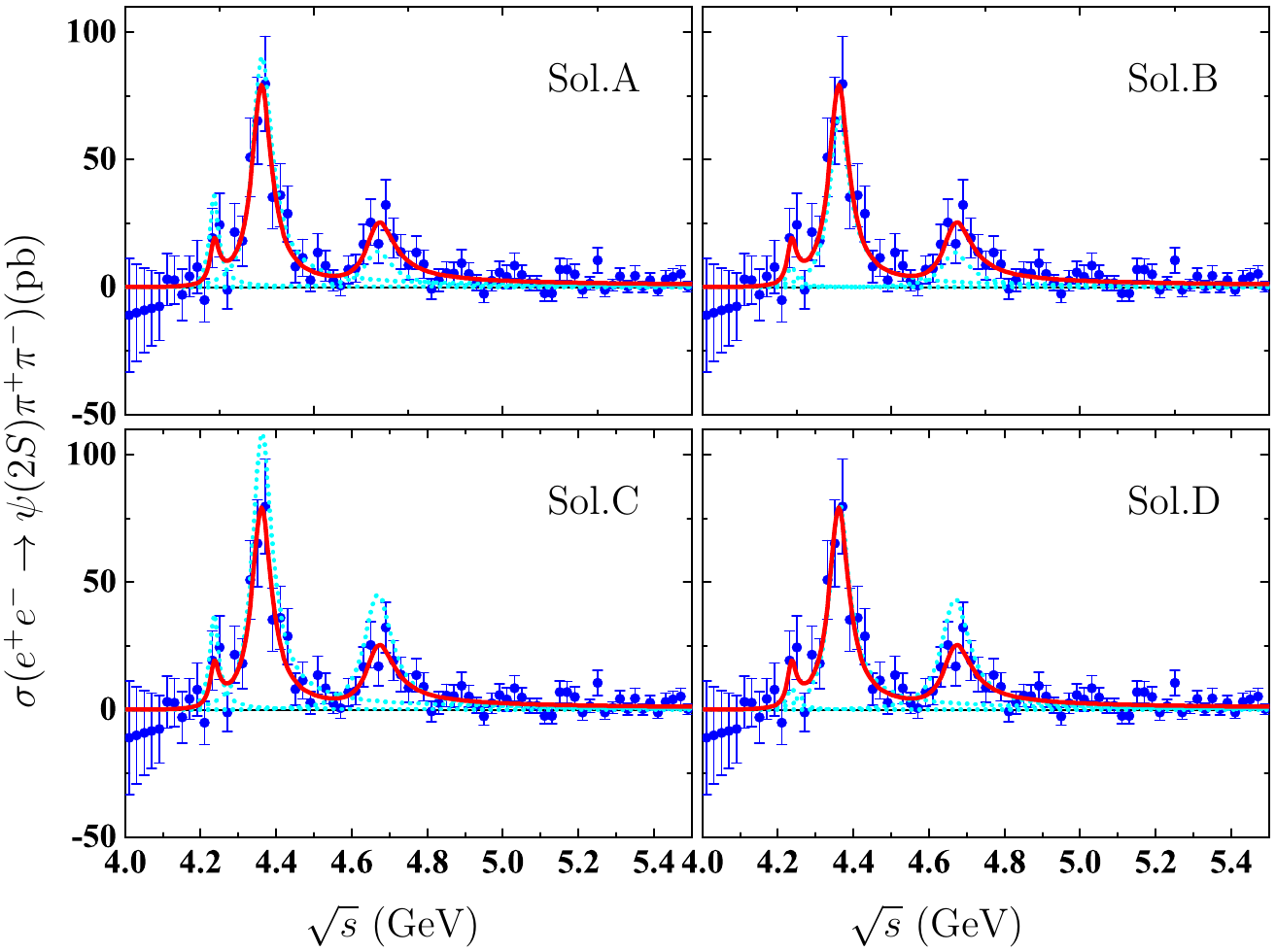}}
\caption{(color online). The different solutions of the resonance contributions and our fitting results for the cross section for $e^+ e^- \to \psi(2S) \pi^+ \pi^-$ in Scheme I. The cyan dashed and red solid curves are the resonance contributions and the fitting results, respectively.  \label{Fig:Sch1-psi2s}}
\end{figure}

\begin{figure}[htb]
\centering%
\scalebox{0.5}{\includegraphics{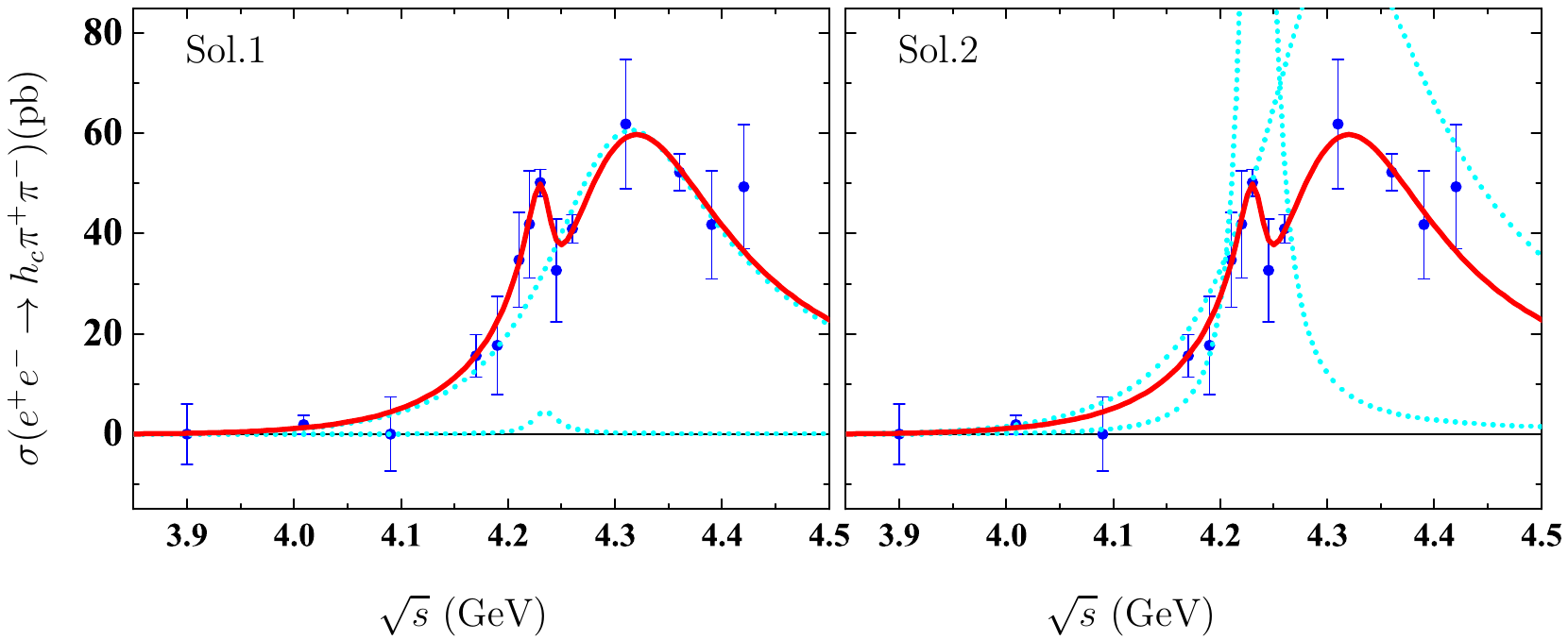}}
\caption{(color online). The different solutions of the resonance contributions and our fitting results for the cross section for $e^+ e^- \to h_c \pi^+ \pi^-$ in Scheme I. The cyan dashed and red solid curves are the resonance contributions and the fitting results, respectively.  \label{Fig:Sch1-hc}}
\end{figure}

\begin{figure}[htb]
\centering%
\scalebox{0.3}{\includegraphics{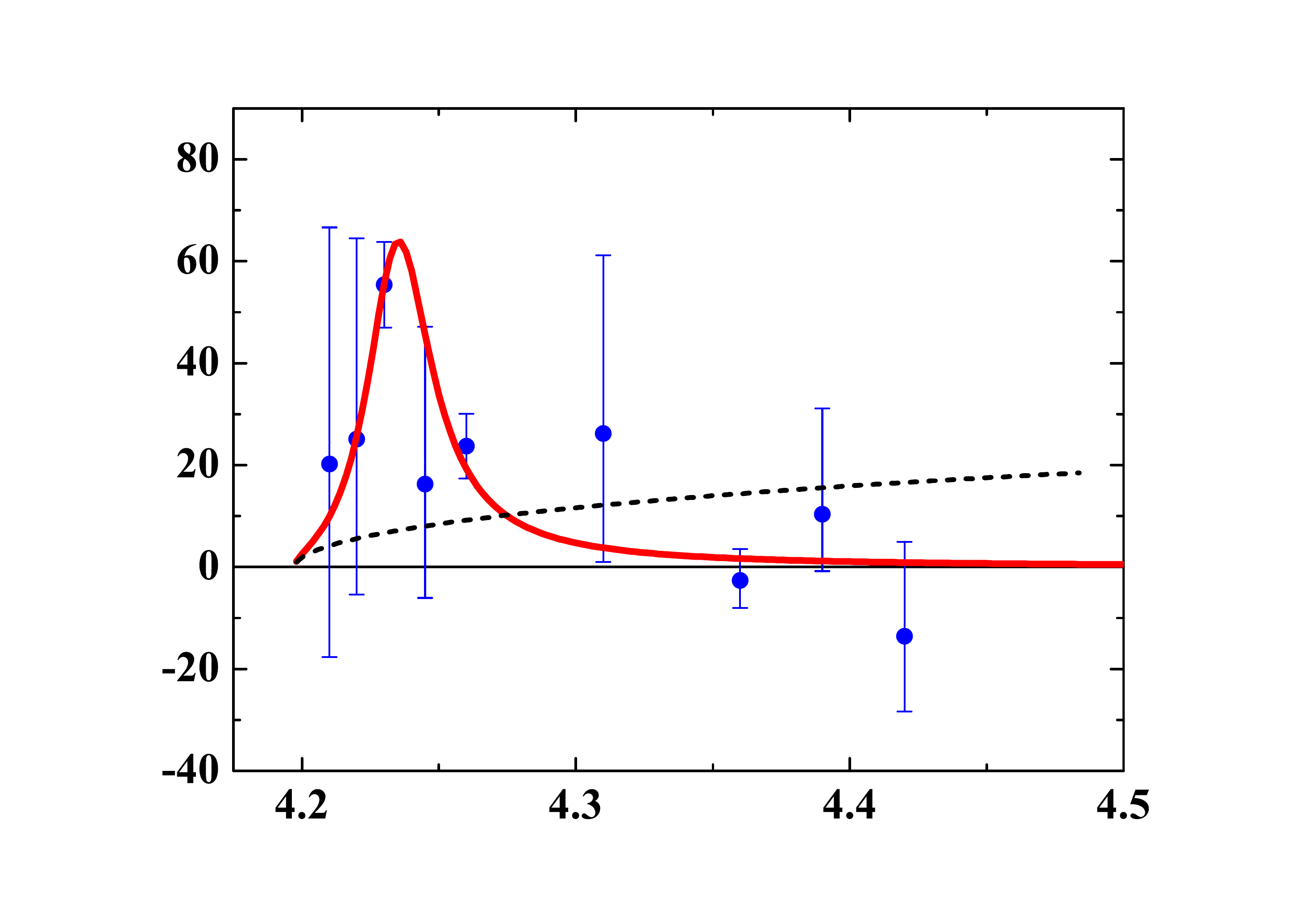}}
\caption{(color online). The different solutions of the resonance contributions and our fitting results for the cross section for $e^+ e^- \to \chi_{c0} \omega$ (solid curve) in Scheme I. The dashed curve is the phase space of $e^+ e^- \to \chi_{c0} \omega$.  \label{Fig:Sch1-chic0}}
\end{figure}

\begin{table*}
\centering %
\caption{The parameters determined by fitting the experimental data of $e^+ e^- \to \psi(2S) \pi^+ \pi^-,\  h_c \pi^+ \pi^-,\ \chi_{c0} \omega$ simultaneously, where the experimental data of $e^+ e^- \to h_c \pi^ + \pi^-$ are depicted by two Breit-Wigner structures.  The masses and the total decay widths are in units of MeV, while the product of the branching ratios are in units of eV. \label{Tab:CombineFit-I}}
\begin{tabular}{cccccccc}
\toprule[1pt]
final State   &  \multicolumn{4}{c} {$\psi(2S) \pi^+ \pi^-$}  & \multicolumn{2}{c}{$h_c \pi^+ \pi^-$}  & $\chi_{c0} \omega$\\
              &  Sol. A & Sol. B & Sol. C & Sol. D & Sol. 1 & Sol. 2 & \\
\midrule[1pt]
$m_{Y(4230)}    $   & \multicolumn{7}{c}{  $ 4234   \pm 5 $}    \\
$\Gamma_{Y(4230)}  $   & \multicolumn{7}{c}{  $ 29  \pm 14 $}   \\
$\Gamma_{Y(4230)}^{e^+e^-} \mathcal{B}(\psi(4S) \to f) $   &   $ 1.3  \pm 0.5  $ & $0.3 \pm 0.2$ & $1.3 \pm 0.5$ & $0.3 \pm 0.3$ & $0.2  \pm 0.1 $ & $7.1 \pm 2.9 $ & $2.2 \pm  0.6$\\
\midrule[1pt]
$m_{Y(4300)}    $   &   $\cdots$  &   $\cdots$ &   $\cdots$ &   $\cdots$  &  \multicolumn{2}{c}{$4294 \pm 11$} & $\cdots$   \\
$\Gamma_{Y(4300)}    $   &   $\cdots$  &   $\cdots$ &   $\cdots$ &   $\cdots$  &  \multicolumn{2}{c}{$201 \pm 55$}& $\cdots$   \\
$\Gamma_{Y(4300)}^{e^+ e^-} \mathcal{B}(Y(4300)\to f)  $   &   $\cdots$  &   $\cdots$ &   $\cdots$ &   $\cdots$  & $14.7 \pm  2.0$ & $23.9 \pm 2.4 $ & $\cdots$ \\
$\phi_1   $   &  $\cdots$  &   $\cdots$ &   $\cdots$ &   $\cdots$ & $5.7 \pm 0.8$ & $3.7 \pm 0.1$& $\cdots$  \\
\midrule[1pt]
$m_{Y(4360)}    $    &  \multicolumn{4}{c}{ $4359 \pm 7$} &$\cdots$ &$\cdots$& $\cdots$  \\
$\Gamma_{Y(4360)}$   &  \multicolumn{4}{c}{$64  \pm 11$} &$\cdots$ &$\cdots$& $\cdots$ \\
$\Gamma_{Y(4360)}^{e^+e^-} \mathcal{B}(Y(4360)\to f)  $   &   $ 7.4  \pm 1.4 $ & $5.5 \pm 1.9 $& $8.9 \pm 1.0$ & $6.6 \pm 1.0$ &$\cdots$ &$\cdots$& $\cdots$ \\
$\phi_2   $   &   $ 4.2  \pm 0.4    $ & $1.5 \pm 0.9$& $4.4 \pm 0.4$& $1.7 \pm 0.6 $ &$\cdots$ &$\cdots$& $\cdots$ \\
\midrule[1pt]
$m_{Y(4660)}    $   &   \multicolumn{4}{c}{$ 4666 \pm 28 $} &$\cdots$ &$\cdots$& $\cdots$   \\
$\Gamma_{Y(4660)}    $   & \multicolumn{4}{c}{ $90  \pm 20 $ } &$\cdots$  &$\cdots$& $\cdots$   \\
$\Gamma_{Y(4660)}^{e^+ e^-} \mathcal{B}(Y(4660)\to f)  $   &   $ 1.9  \pm 0.8     $   &$1.8 \pm 0.7$& $ 6.0 \pm 3.2 $ &$5.8 \pm 2.3$ &$\cdots$&$\cdots$ & $\cdots$ \\
$\phi_3   $   &   $ 5.2   \pm 0.7    $  &$2.2 \pm 1.0$& $3.1 \pm 0.5$& $ 0.1 \pm 2.1 $ &$\cdots$&$\cdots$ & $\cdots$ \\
\midrule[1pt]
$\chi^2/\mathrm{ndf}$ &  \multicolumn{7}{c}{52.2/81}\\
\bottomrule[1pt]

\end{tabular}
\end{table*}
\begin{table*}[htb]
\centering
\caption{Our estimate of the ratios of the branching ratios of the $Y(4230)$ dipion transitions to the one of $Y(4230) \to \chi_{c0} \omega$. \label{Tab:Ratio}}
\begin{tabular}{c|cccc|cc}
\toprule[1pt]
              & \multicolumn{4}{c|}{$\mathcal{R}^{\psi(2S) \pi^+ \pi^-}_{\chi_{c0} \omega}$ } & \multicolumn{2}{c}{$\mathcal{R}^{h_c \pi^+ \pi^-}_{\chi_{c0} \omega}$}\\
\cline{2-7}
             & Sol. A & Sol. B & Sol. C & Sol. D & Sol. 1 & Sol.2 \\
\midrule[1pt]
Scheme I  & $0.56 \pm 0.25$ & $0.13 \pm 0.12$ & $0.59 \pm 0.26$ & $0.14 \pm 0.13$  & $0.07 \pm 0.05$ & $3.21 \pm 0.97$ \\
Scheme II & $0.06 \pm 0.09$ & $0.26 \pm 0.30$ &$0.27 \pm 0.20$ & $ 0.05 \pm 0.08$& \multicolumn{2}{c}{$0.11 \pm 0.04$}\\
\bottomrule[1pt]
\end{tabular}
\end{table*}

In this scheme, we simulate the cross section for $e^+e^- \to \psi(2S) \pi^+ \pi^-$ , $e^+  e^- \to h_c \pi^+ \pi^-$  and $e^+ e^- \to \chi_{c0} \omega$ by three resonances, two resonances and one resonance, respectively. A resonance near 4.2 GeV, $Y(4230)$, is included in all three processes. The parameters determined by fitting the experimental data are listed in Table \ref{Tab:CombineFit-I}. With these parameters , the $\chi^2/\mathrm{ndf}$ is estimated to be $52.2/81$. The resonance parameters of the $Y(4230)$ included in all three processes are fitted to be
\begin{equation}\label{eq:psi4S2}
\begin{split}
m_{Y(4230)} &= 4234 \pm 5 \ \mathrm{MeV}, \\
\Gamma_{Y(4230)} &= 29 \pm 14\ \mathrm{MeV} .
\end{split}
\end{equation}
The resonance parameters of $Y(4230)$ obtained by a combined fit are consistent with those determined by fitting the cross sections for $e^+ e^- \to h_c \pi^+ \pi^-$ and $e^+ e^- \to \chi_{c0} \omega$  \cite{Ablikim:2014qwy, Yuan:2013ffw} separately. In addition, the narrow width obtained is also consistent with our expectation of $\psi(4S)$ in Ref. \cite{He:2014xna}.

Besides $Y(4230)$, two other resonances $Y(4360)$ and $Y(4660)$ are also involved in the cross section for $e^+ e^- \to \psi(2S) \pi^+ \pi^-$. The resonance parameters of these two charmonium-like states are determined to be $m_{Y(4360)}=4359 \pm 7 \ \mathrm{MeV},\ \Gamma_{Y(4360)}=64 \pm 11 \ \mathrm{MeV}, \ m_{Y(4660)}=4666 \pm 28\ \mathrm{MeV}, \ \Gamma_{Y(4660)}= 90 \pm 20\ \mathrm{MeV} $, respectively,
which are consistent with the corresponding PDG average values \cite{Agashe:2014kda}. When simulating the cross section for $e^+ e^- \to \psi(2S) \pi^+ \pi^-$ by three Breit-Wigner functions, we get four different solutions with equally good fit quality, whose parameters for different solutions are listed in Table \ref{Tab:CombineFit-I}. Among these four solutions, the masses and widths of $Y(4230)$, $Y(4360)$ and $Y(4660)$ are the same, but the product $\Gamma_{R}^{e^+e^-} \times \mathcal{B}_{R\to \psi(2S) \pi^+ \pi^-}$ with $R=(Y(4230), Y(4360), Y(4660))$ and the phase angles $\phi_i$ are different. For Sol. A and Sol. C, interference between $Y(4230)$ and $Y(4360)$ is destructive and $\Gamma \mathcal{B}=1.3 \pm 0.5$ for $Y(4230)$, while for Sol. B and Sol. D, interference is constructive.
In Fig. \ref{Fig:Sch1-psi2s}, the resonance contributions from $Y(4230)$, $Y(4360)$ and $Y(4660)$ are presented for different solutions. For Sol. A and Sol. C, the contribution from $Y(4230)$ is significant, while for Sol. B and Sol. D,
it is relatively small.

As for the cross section for $e^+ e^- \to h_c \pi^+ \pi^-$, it is simulated by two Breit-Wigner functions for $Y(4230)$ and $Y(4300)$. In contrast to $Y(4230)$, the charmonium-like state $Y(4300)$ is a very broad structure and its resonance parameters are $m_{Y(4300)}=4294 \pm 11$ MeV and $\Gamma_{Y(4300)}= 201 \pm 55$ MeV. Similarly to the case of $e^+ e^- \to \psi(2S) \pi^+ \pi^-$, there exist two different solutions to the cross section for $e^+ e^- \to h_c \pi^+ \pi^-$. {In Fig. \ref{Fig:Sch1-hc}, the fitting results and the resonance contributions are presented. Comparing the two solutions, we conclude that these two resonances interfere constructively and destructively for Sol. 1 and Sol. 2, respectively}. The cross section for $e^+ e^- \to \chi_{c0} \omega$ is simulated by one resonance $Y(4230)$, whose fitting results are presented in Fig. \ref{Fig:Sch1-chic0}. Since the experimental data for this process are not abundant and have large errors, we also use the phase space of $e^+ e^- \to \chi_{c0} \omega$ to fit the experimental data. The fitting curve is presented in Fig. \ref{Fig:Sch1-chic0} with the $\chi^2$ of 49.8, while the $\chi^2$ of fitting the data with a resonance is $3.1$. In addition, we further check the possibility of fitting the experimental data with the sum of the phase space and the resonance; the fitted $\chi^2$ is almost the same as that of fitting the data only with a resonance and hence the contribution from the phase space is ignorable, which indicates that fitting the cross sections of $e^+ e^- \to \chi_{c0} \omega$ with a resonance is more reasonable. On the experimental side, we expect that additional measurements from BESIII and the forthcoming BelleII for this process will provide a further restriction on the properties of $Y(4230)$.

In the combined fit, we include the contributions of $Y(4230)$ in all three of these hidden charm production channels. Thus we can compare the branching ratios of $Y(4230)$ decaying into $\psi(2S) \pi^+ \pi^-$, $h_c\pi^+ \pi^-$, and $\chi_{c0} \omega$.  From our combined fit, we find the product branching ratio for the $\chi_{c0} \omega $ channel is $\Gamma_{Y(4230)}^{e^+ e^-} \mathcal{B}(Y(4230) \to \chi_{c0} \omega)=2.2 \pm 0.6$ eV, which is consistent with the value, $2.7 \pm 0.4 \pm 0.5$ eV, reported by the BESIII Collaboration \cite{Ablikim:2014qwy}. We list the ratios of $\mathcal{B}(Y(4230) \to \psi(2S) \pi^ +\pi^-)$ and $\mathcal{B}(Y(4230)\to h_c \pi^+ \pi)$ to $\mathcal{B}(Y(4230) \to \chi_{c0} \omega)$ in Table \ref{Tab:Ratio}. In Scheme I, the ratio of $\mathcal{B}(Y(4230) \to \psi(2S) \pi^+ \pi^-)$ to  $\mathcal{B}(Y(4230) \to \chi_{c0} \omega)$ is estimated to be $0.56 \pm 0.25$ and $0.59 \pm 0.26$ for Sol. A and Sol. B, respectively, in which interference between $Y(4230)$ and $Y(4360)$ is destructive as shown in Fig. \ref{Fig:Sch1-psi2s}. If interference is constructive, the ratio $\mathcal{R}^{\psi(2S) \pi^+ \pi^-}_{\chi_{c0} \omega}$ is relatively as small as $0.13 \pm 0.12$ and $0.14 \pm 0.13$ for Sol. B and Sol. D, respectively.

As for the ratio of $\mathcal{B}(Y(4230) \to h_c \pi^+ \pi^-)$ to $\mathcal{B}(Y(4230) \to \chi_{c0}\omega)$, it is obtained as $0.07 \pm 0.05$ and $3.21 \pm 0.97$ for Sol. 1 and Sol. 2, in which the interference between $Y(4230)$ and $Y(4300)$ is constructive and destructive, respectively. In addition, if $Y(4230)$ is the $\psi(4S)$, the spins of charm and anti-charm quarks in $\psi(4S)$ are parallel, while in $h_c$ they are antiparallel. Thus, the $\psi(4S) \to h_c \pi^+ \pi^-$ decay is a spin flip process, which should be suppressed in heavy quark effective theory. Thus, the ratio $\mathcal{R}^{h_c \pi^+ \pi^-}_{\chi_{c0}\omega}$ for Sol. 1, i.e., $0.07 \pm 0.05$, is more favored as a physical solution compared with the one for Sol. 2.

\subsection{Scheme II}

\begin{table*}
\centering %
\caption{The same as Table \ref{Tab:CombineFit-I} but in Scheme II. \label{Tab:CombineFit-II}}
\begin{tabular}{ccccccc}
\toprule[1pt]
final State   &  \multicolumn{4}{c} {$\psi(2S) \pi^+ \pi^-$}  & $h_c \pi^+ \pi^-$  & $\chi_{c0} \omega$\\
              &  Sol. a & Sol. b & Sol. c & Sol. d &  & \\
\midrule[1pt]
$m_{Y(4230)}    $   & \multicolumn{6}{c}{  $ 4220   \pm 8 $}    \\
$\Gamma_{Y(4230)}  $   & \multicolumn{6}{c}{  $ 43  \pm 9 $}   \\
$\Gamma_{Y(4230)}^{e^+e^-} \mathcal{B}(Y(4230) \to f) $   &   $ 0.2 \pm 0.3 $ & $0.7 \pm 1.1$ & $0.8 \pm 0.6$ & $0.3 \pm 0.3$ & $0.3  \pm 0.2 $  & $2.8 \pm  0.9$\\
\midrule[1pt]
$m_{Y(4360)}    $    &  \multicolumn{4}{c}{ $4360 \pm 7$} &$\cdots$ & $\cdots$  \\
$\Gamma_{Y(4360)}$   &  \multicolumn{4}{c}{$68  \pm 14$} &$\cdots$ & $\cdots$ \\
$\Gamma_{Y(4360)}^{e^+e^-} \mathcal{B}(Y(4360)\to f)  $   &   $ 7.5  \pm 1.1 $ & $7.3 \pm 1.5 $& $9.0 \pm 1.2$ & $6.1 \pm 2.2$ &$\cdots$& $\cdots$ \\
$\phi_2   $   &   $ 2.7  \pm 1.9    $ & $3.9 \pm 0.7$& $4.1 \pm 0.7$& $2.4 \pm 1.9 $ &$\cdots$& $\cdots$ \\
\midrule[1pt]
$m_{Y(4660)}    $   &   \multicolumn{4}{c}{$ 4664 \pm 27 $}  &$\cdots$& $\cdots$   \\
$\Gamma_{Y(4660)}    $   & \multicolumn{4}{c}{ $93  \pm 21 $ }  &$\cdots$& $\cdots$   \\
$\Gamma_{Y(4660)}^{e^+ e^-} \mathcal{B}(Y(4660)\to f)  $   &   $ 5.9  \pm 2.4     $   &$2.0 \pm 0.6$& $ 6.1 \pm 1.0 $ &$1.9 \pm 0.9$ &$\cdots$ & $\cdots$ \\
$\phi_3   $   &   $ 1.0   \pm 1.9    $  &$4.8 \pm 1.1$& $2.8 \pm 1.0$& $ 3.1 \pm 2.0 $ &$\cdots$ & $\cdots$ \\
\midrule[1pt]
$c_0$ &$\cdots$ &$\cdots$ &$\cdots$ &$\cdots$ & $12279 \pm 551$ & $\cdots$\\
$\phi_4$ &$\cdots$ &$\cdots$ &$\cdots$ &$\cdots$ & $5.6 \pm 0.4$ & $\cdots$\\
\midrule[1pt]
$\chi^2/\mathrm{ndf}$ &  \multicolumn{6}{c}{63.0/83}\\
\bottomrule[1pt]

\end{tabular}
\end{table*}

\begin{figure}[htb]
\centering%
\scalebox{0.63}{\includegraphics{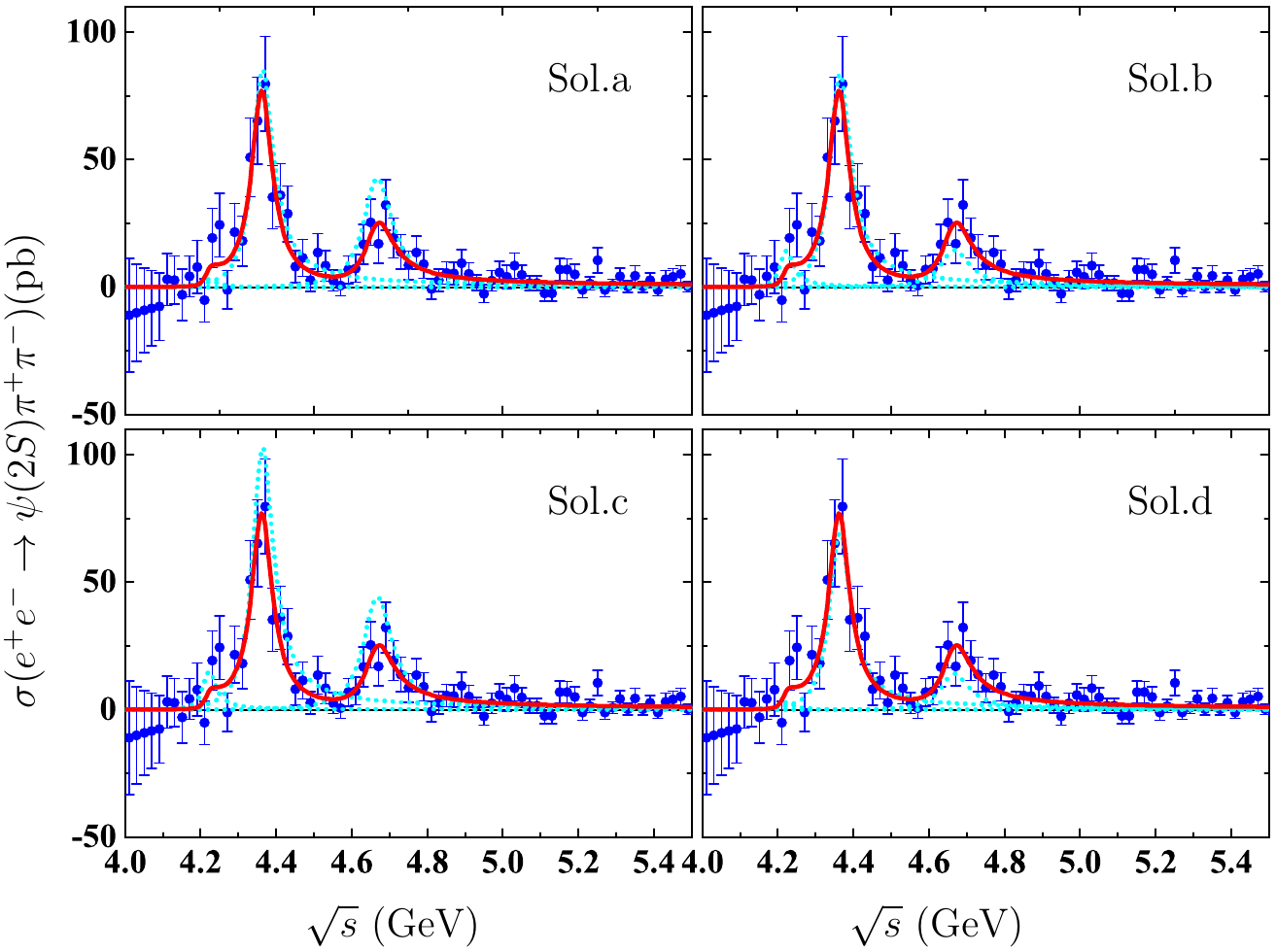}}
\caption{(color online). The same as Fig. \ref{Fig:Sch1-psi2s} but in scheme II. \label{Fig:Sch2-psi2s}}
\end{figure}

\begin{figure}[htb]
\centering%
\scalebox{0.68}{\includegraphics{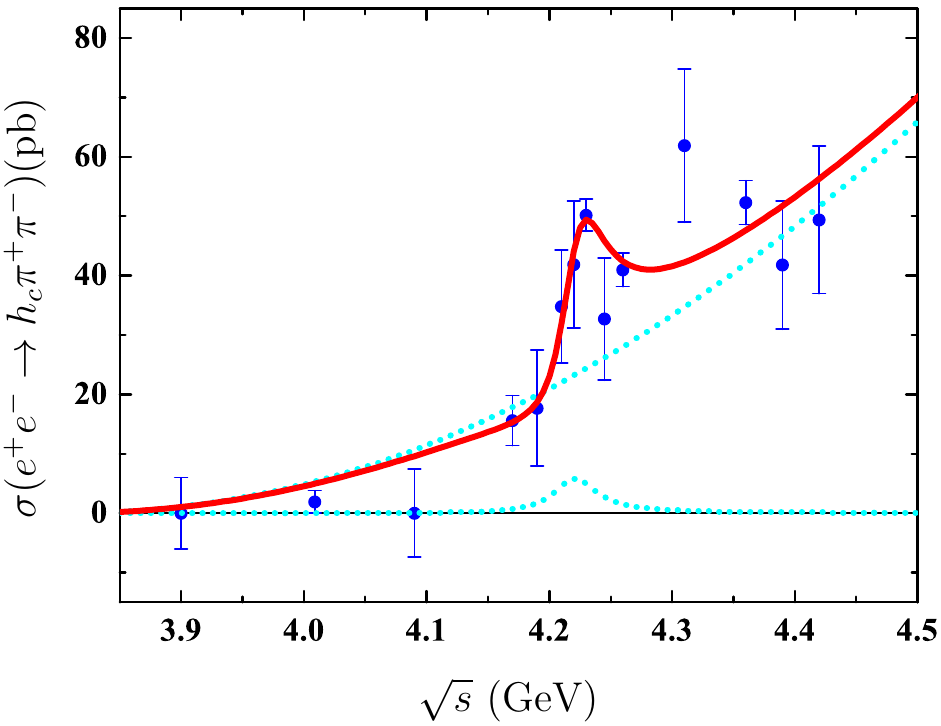}}
\caption{(color online). The same as Fig. \ref{Fig:Sch1-hc} but in scheme II. \label{Fig:Sch2-hc}}
\end{figure}

\begin{figure}[htb]
\centering%
\scalebox{0.65}{\includegraphics{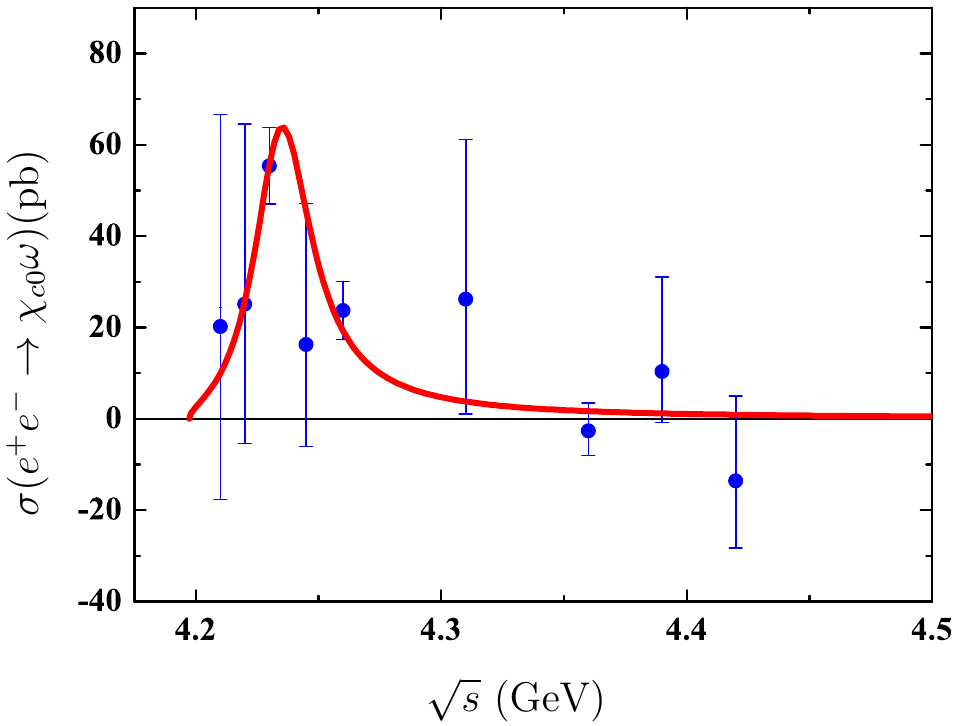}}
\caption{(color online). The same as Fig. \ref{Fig:Sch1-chic0} but in scheme II. \label{Fig:Sch2-chic0}}
\end{figure}

In Scheme II, the cross section for $e^+ e^- \to h_c \pi^+ \pi^-$ is simulated by a coherent sum of a Breit-Wigner function and $2\to 3$ phase space as shown in Eq.~(\ref{Eq:hcII}), in which the cross section for $e^+ e^- \to h_c \pi^+ \pi^-$ goes up as the energy of the center-of-mass system is increased. The cross sections for $e^+ e^- \to \psi(2S) \pi^+ \pi^-$ and $e^+ e^- \to \chi_{c0} \omega$ are depicted by three Breit-Winger functions and one Breit-Winger function, respectively, which are the same as those in Scheme I. The fitted parameters are presented in Table \ref{Tab:CombineFit-II}, with which the $\chi^2 /\mathrm{ndf}$ is estimated to be $63.0/83$ and is a bit larger than the one in Scheme I. In Scheme II, the resonance parameters of the $Y(4230)$ are fitted to be
\begin{equation}\label{eq:psi4S3}
\begin{split}
m_{Y(4230)} &= 4220 \pm 8 \ \mathrm{MeV}, \\
\Gamma_{Y(4230)}&=  43 \pm 9 \ \mathrm{MeV}.
\end{split}
\end{equation}
The center value of the $Y(4230)$ mass fitted in Scheme II is smaller than the one obtained in Scheme I, while the center value of the total width in Scheme II is much larger than Scheme I.

Sine $Y(4360)$ and $Y(4660)$ are not involved in the other two processes, the resonance parameters of these two states are determined as $m_{Y(4360)}=4360 \pm 7 \ \mathrm{MeV}, \ \Gamma_{Y(4360)}=68 \pm 14\ \mathrm{MeV}, \ m_{Y(4360)}=4664 \pm 27\ \mathrm{MeV}$, and $\Gamma_{Y(4360)}=93 \pm 21 \ \mathrm{MeV}$, respectively, by fitting the cross section for $e^+ e^- \to \psi(2S) \pi^+\pi^-$ in Scheme II, which is consistent with the parameters in Scheme I within errors. The four solutions $a-d$ to the cross section for $e^+ e^- \to \psi(2S) \pi^+ \pi^-$ in Scheme II are presented in Fig. \ref{Fig:Sch2-psi2s}. One finds that the signal of $Y(4230)$ is inconspicuous in the cross section for $e^+ e^- \to \psi(2S) \pi^+ \pi^-$, which is caused by the relative large total width of $Y(4230)$ fitted in Scheme II. In this case, the product $\Gamma_{Y(4230) }^{ e^+ e^-}  \mathcal{B}(Y(4230) \to \psi(2S) \pi^+ \pi^-)$ is fitted to be $0.2 \pm 0.3$, $0. 7 \pm 1.1$, $0.8 \pm 0.6$ and $0.3 \pm 0.3$ for the four solutions, respectively. All these fitted products $\Gamma_{Y(4230) }^{ e^+ e^-}  \mathcal{B}(Y(4230) \to \psi(2S) \pi^+ \pi^-)$ have large errors, which indicates that the data of the cross section for $e^+ e^- \to \psi(2S) \pi^+\pi^-$ are in disfavor with the broad $Y(4230)$.

The fitting results and the resonance contributions to the cross section for $e^+ e^- \to h_c \pi^+ \pi^-$ are presented in Fig. \ref{Fig:Sch2-hc}. The cross section goes up with the increase of the energy of the center-of-mass system above 4.4 GeV and behaves as the phase space of $e^+e^- \to h_c \pi^+ \pi^-$. Further experimental measurements of this cross section, especially above 4.4 GeV, will show the explicit tendency of the cross section, which will provide a crucial test for both Schemes. The cross section for $e^+ e^- \to \chi_{c0} \omega$ in Scheme II is presented in Fig. \ref{Fig:Sch2-chic0}. The $Y(4230)$ with a lower mass and wide width also provides a good description for this cross section due to the large error of the experimental data.

As in Scheme I,
the ratios of $\mathcal{B}(Y(4230) \to \psi(2S) \pi^+\pi^-)$ and $\mathcal{B}(Y(4230) \to h_c \pi^+\pi^-)$ to $\mathcal{B}(Y(4230) \to \chi_{c0} \omega)$ can be estimated by performing the combined fit and are listed In Table \ref{Tab:Ratio}. The ratio $R^{\psi(2S) \pi^+\pi^-}_{ \chi_{c0} \omega}$ has a large error and is compatible with zero. The large error dominantly comes from the error of the branching ratio of $Y(4230) \to \psi(2S) \pi^+\pi^-$. As for the ratio $R^{h_c \pi^+\pi^-}_{ \chi_{c0} \omega}$, it is fitted to be $0.11 \pm 0.04$, which is consistent with the expectation of heavy quark effective theory when taking $Y(4230)$ as $\psi(4S)$.


\section{Summary} \label{sec5}
Being stimulated by the anomalous mass gaps of the $S$-wave charmonia and the similarity between the charmonium and bottomonium families, we have predicted a missing $S$-wave vector charmonium state, $\psi(4S)$, located near $4.2$ GeV in Ref. \cite{He:2014xna}. Theoretical estimates based on the quark pair creation model indicate that this predicted $\psi(4S)$ is a narrow state \cite{He:2014xna}. Since on the experimental side a narrow structure near 4.2 GeV has been observed in the $e^+ e^- \to h_c \pi^+ \pi^-$ \cite{Yuan:2013ffw} and $e^+ e^- \to \chi_{c0} \omega$ processes \cite{Wang:2014hta}, we have attributed the structure in both processes to the missing $\psi(4S)$. The search for the signal of $\psi(4S)$ in other channels is an intriguing and urgent problem.

Recently, more precise experimental measurements of the cross section for $e^+ e^- \to \psi(2S) \pi^+ \pi^-$ have been performed by the Belle Collaboration \cite{Wang:2014hta}. Since other than the structures of $Y(4360)$ and $Y(4660)$, a number of bump events near 4.2 GeV have been found in the same cross section, in this paper we have performed a fit to the cross section with three resonances to find one additional resonance, $Y(4230)$, with $m=4243$ MeV and $\Gamma=16 \pm 31$ MeV. The resonance parameters of the $Y(4230)$ are consistent with our expectation for the missing $\psi(4S)$. The upper limit of the  product $\Gamma_{Y(4230)}^{e^+ e^-} \mathcal{B}(Y(4230) \to \psi(2S) \pi^+ \pi^-)$ is fitted to be $1.9 $ eV. With the assumption of $Y(4230)$ as the $\psi(4S)$ and the theoretical estimate of $\Gamma_{\psi(4S)}^{e^+ e^-}$ in a screening potential model, the branching ratio of $\psi(4S) \to \psi(2S) \pi^+ \pi^-$ is evaluated to be $\mathcal{B}(\psi(4S) \to \psi(2S) \pi^+ \pi^-)< 3\times 10^{-3}$.

One should notice that this mass of the $\psi(4S)$ is above the threshold of a charmed or charmed-strange meson pair, which indicates that the dominant decay modes of $\psi(4S)$ are open-charm decays. As for the hidden charm decay processes, like $\psi(4S) \to \psi(2S) \pi^+ \pi^-$ and $\chi_{c0} \omega$, they can occur via charmed or charm-stranged meson loops. We have calculated the meson loop contributions to the decay $\psi(4S) \to \psi(2S) \pi^+ \pi^-$ to find that the branching ratio resulting from the meson loops is dependent on the parameters involved in calculations, but the branching ratio from our fit to the cross section for $e^+ e^- \to \psi(2S) \pi^+ \pi^-$ is understandable in a reasonable parameter range, which also indicates that the $Y(4230)$ observed in the $e^+ e^- \to \psi(2S) \pi^+ \pi^-$ process could be a good candidate of the $\psi(4S)$.

To further test our conjecture that the structures near 4.2 GeV in $e^+ e^- \to h_c  \pi^+ \pi^-$, $e^+ e^- \to \psi(2S) \pi^+ \pi^-$, and $e^+ e^- \to \chi_{c0} \omega$ come from the same source, i.e., $\psi(4S)$, we have performed a combined fit to these three hidden charm production processes in two different schemes. The mass and width of $\psi(4S)$ are fitting to be $m_{\psi(4S)}=4234 \pm 5$ MeV, $\Gamma_{\psi(4S)} =29 \pm 14$ MeV and $m_{\psi(4S)}=4220 \pm 8$ MeV, $\Gamma_{\psi(4S)} =43 \pm 9$ MeV for Scheme I and Scheme II, respectively, as shown in Eqs. (\ref{eq:psi4S2}) and (\ref{eq:psi4S3}). Besides the resonance parameters of $\psi(4S)$, the products of the branching ratios of $\psi(4S) \to \psi(2S) \pi^+ \pi^-,\ h_c\pi^+ \pi^-, \ \chi_{c0} \omega$ and the dilepton decay width of the $\psi(4S)$ are also determined by fitting the experimental data. The ratios of the branching ratios for the  hidden dipion decay to the $\chi_{c0} \omega$ decay mode are also estimated, which can be tested by further experimental measurements at BESIII and the forthcoming BelleII.

Before closing this section, we would like to emphasize that the precise experimental measurements of the hidden-charm decay have provided some evidences of the existence of the missing $\psi(4S)$ at the present moment. The future measurement for the hidden-charm decay will be able to further test our estimate of resonance parameters of $\psi(4S)$, in particular, the measurements of the cross section for $e^+ e^- \to h_c \pi^+ \pi^-$ above 4.4 GeV can provide a further restriction on the properties of $\psi(4S)$. In addition, the precise measurements for the open-charm decay mode are also crucial for the properties of the missing $\psi(4S)$. We hope that precise measurements for both open-charm and hidden-charm decay modes will be carried out at BESIII and the forthcoming BelleII.

\section*{Acknowledgments}
We would like to thank Cheng-Ping Shen for his useful discussions. This project is in part supported by the National Natural Science Foundation
of China under Grants No. 11222547, No. 11375240, No. 11175073, and
No. 11035006, and the Ministry of Education of China (SRFDP under Grant
No. 2012021111000).

\end{document}